\documentstyle[12pt]{article}
\begin{document}
\begin{center}
{\Large
\bf Ground state hyperfine splitting of high Z hydrogenlike ions}\\
\end{center}

\begin{center}
{V.M.Shabaev$^{1}$, M.Tomaselli$^{2}$, T.K\"uhl$^{3}$,
A.N.Artemyev$^{1}$, and V.A.Yerokhin $^{1}$}
\end{center}

$^{1}$ {\it Department of Physics, St.Petersburg
State University,\\ Oulianovskaya Street 1, Petrodvorets,
St.Petersburg 198904, Russia}\\
$^{2}$ {\it Institute f\"ur Kernphysik, Technische Hochschule Darmstadt,
Schlossgartenstrasse 9, D-64289 Darmstadt, Germany}\\
$^{3}$ {\it Gesellschaft f\"ur Schwerionenforschung (GSI), Postfach 11 05 52,
D-64223 Darmstadt, Germany}

\begin{abstract}
The ground state hyperfine splitting values of high Z hydrogenlike
ions are calculated. The relativistic, nuclear and QED corrections
are taken into account. The nuclear magnetization distribution
correction (the Bohr-Weisskopf effect) is evaluated within the
single particle model with the $g_{S}$-factor chosen to yield
the observed nuclear moment. An additional contribution
caused by the nuclear spin-orbit interaction is included in the
calculation of the Bohr-Weisskopf effect.
 It is found that the theoretical
value of the wavelength of the transition between the hyperfine
splitting components in $^{165}Ho^{66+}$
is in good agreement with experiment.
\\
$\;$
\\
PACS number(s): 31.30.Gs, 31.30.Jv
\end{abstract}

\newpage
\section{Introduction}
Laser spectroscopic measurement of the ground state hyperfine splitting
in hydrogenlike  $^{209}Bi^{82+}$  [1] has
triggered great interest to calculations of this value
(see [2-7] and references therein).
Recently the transition  between the  $F=4$ and $F=3$
 hyperfine splitting levels of the ground state in hydrogenlike
$^{165}Ho^{66+}$ was observed [8] and its
wavelength was determined to be $572.79(15)$ nm. It was found [8]
that this value is in disagreement with commonly tabulated values of the
nuclear dipole magnetic moment of $^{165}Ho$ (see, e.g., [9]) and
 in good  agreement with
the value that was measured by Nachtsheim [10] and compiled by Peker [11].

The hyperfine splitting values of $^{165}Ho^{66+}$
(for the magnetic moment from [9])
and other possible candidates for such experiments were evaluated,
without QED corrections, in [4]. In the present paper we refine
the calculations of [4] considering
a more accurate treatment of the
nuclear effects and taking into account the QED corrections.

\section{Basic formulas and calculations}

The ground state hyperfine splitting  of hydrogenlike ions
is conveniently written in the form [4,12]
\begin{eqnarray}
\Delta E_{\mu}&=&\frac{4}{3}\alpha(\alpha Z)^{3}\frac{\mu}{\mu_{N}}
\frac{m}{m_{p}}\frac{2I+1}{2I}
mc^{2}\nonumber\\
& &\times\{A(\alpha Z)(1-\delta)(1-\varepsilon)+x_{rad}\}\;.
\end{eqnarray}
Here $\alpha$ is the fine structure constant, $ Z $ is the nuclear
charge, $m$ is the electron mass,
 $m_{p}$ is the proton mass,
$\mu$ is the nuclear magnetic moment, $\mu_{N}$ is the nuclear
magneton, $I$ is the nuclear spin.
 $A(\alpha Z)$ denotes
the relativistic factor [13]
\begin{eqnarray}
A(\alpha Z)=\frac{1}{\gamma(2\gamma-1)}=1+\frac{3}{2}(\alpha Z)^{2}
+\frac{17}{8}(\alpha Z)^{4}+\cdots\,\,,
\end{eqnarray}
where $\gamma=\sqrt{1-(\alpha Z)^{2}}$.
 $\delta$ is the nuclear charge
 distribution
correction,
 $\varepsilon$ is the
nuclear  magnetization distribution correction (the Bohr-Weisskopf correction)
[14],
and $x_{rad}$ is the QED correction.

To calculate the nuclear charge distribution correction $\delta$ we used
the two-parameter Fermi model:
\begin{eqnarray}
\rho{(r)}=\frac{\rho_{0}}{1+\exp{[(r-c)/a]}}
\end{eqnarray}
 The  parameters
$c$,  $a$ and $\langle r^{2}\rangle ^{1/2}$ have been taken from [15].
We found that the error due to an uncertainty
 of the nuclear charge distribution
parameters is much smaller
than an uncertainty
 of the Bohr-Weisskopf effect.
The Bohr-Weisskopf effect was calculated assuming that the magnetizations
can be ascribed to the single particle structure of the nucleus
with the effective $g_{S}$-factor chosen to yield the observed nuclear
magnetic moment.
               The nucleon wave functions were calculated by the
Schr\"odinger equation with the Woods-Saxon potential [16-17]
\begin{eqnarray}
U(r)=V(r)+V_{SO}(r)+V_{Coul}(r)\,,
\end{eqnarray}
where
$$
V(r)=-V_{0}f(r)\,,
$$
$$
V_{SO}=\lambda V_{0}(\hbar/2m_{p}c)^{2}
\mbox{\boldmath $\sigma$}
\cdot{\bf l}\;r^{-1}d f_{SO}(r)/d r\,,
$$
$$
V_{Coul}=
\left(\begin{array}{l}
\alpha (Z-1)(3-r^{2}/R_{0}^{2})/2R_{0}\,,\,\,\,\, r\leq R_{0}\\
\alpha (Z-1)/r\,,\,\;\;\;\;\;\;\,\,\,\;\;\;\;\;\;\;\;r\geq R_{0}
\end{array}\right)\;,\\
$$
$$
f(r)=\Bigl[1+\exp{[(r-R_{0})/a]}\Bigr]^{-1},
$$
$$
f_{SO}(r)=\Bigl[1+\exp{[(r-R_{SO})/a]}\Bigr]^{-1}\,.
$$
In the neutron case,
 the term $V_{Coul}$  must be omitted.
The phenomenological  spin-orbit (SO) interaction, characterized by
the parameter $\lambda$, is much larger than it follows from the Dirac equation.
 In the present paper
we used the potential parameters from [17]:
$R_{0}=1.347 A^{\frac{1}{3}}$ fm,
 $V_{0}=40.6$ MeV,
$R_{SO}=1.280 A^{\frac{1}{3}}$ fm,
$\lambda=31.5$
 for neutron and
$R_{0}=1.275 A^{\frac{1}{3}}$ fm,
 $V_{0}=58.7$ MeV,
$R_{SO}=0.932 A^{\frac{1}{3}}$ fm,
$\lambda=17.8$
 for proton.
Despite these parameters were chosen to give
reasonable binding energies
 for the lead region alone,
we used them at lower $Z$ as well.
We found that the uncertainty of the Bohr-Weisskopf effect
caused by possible changes of these parameters [18,19]
is small in comparison with an expected error due to
deviation from the single particle model.

As is known [20], the SO interaction gives an additional
contribution to the nuclear magnetic moment.
Taking into account  the related term in the hyperfine
splitting theory gives an additional contribution to the
Bohr-Weisskopf effect. If we denote the SO interaction
by
\begin{eqnarray}
V_{SO}(r)=\phi_{SO}(r)({\bf s}\cdot {\bf l})\,,
\end{eqnarray}
 the total expressions for the
Bohr-Weisskopf correction within the single particle
approximation are given by
\begin{eqnarray}
\varepsilon&=&\frac{g_{S}}{g_{I}}\Bigl[\frac{1}{2I}\langle K_{S}
\rangle+\frac{(2I-1)}{8I(I+1)}
\langle K_{S}-K_{L} \rangle \Bigr]\nonumber\\
&&+	\frac{g_{L}}{g_{I}}\Bigl[\frac{(2I-1)}{2I}\langle K_{L}
\rangle +\frac{(2I+1)}{4I(I+1)} \frac{m_{p}}{\hbar ^{2}}
\langle\phi_{SO}r^{2}K_{L}\rangle\Bigr]
\end{eqnarray}
 for $I=L+\frac{1}{2}$ and
\begin{eqnarray}
\varepsilon&=&\frac{g_{S}}{g_{I}}\Bigl[-\frac{1}{2(I+1)}\langle K_{S}
\rangle-\frac{(2I+3)}{8I(I+1)}
\langle K_{S}-K_{L} \rangle \Bigr]\nonumber\\
&&+\frac{g_{L}}{g_{I}}\Bigl[\frac{(2I+3)}{2(I+1)}\langle K_{L}
\rangle -\frac{(2I+1)}{4I(I+1)} \frac{m_{p}}{\hbar ^{2}}\langle
\phi_{SO}r^{2}K_{L}\rangle\Bigr]
\end{eqnarray}
 for $I=L-\frac{1}{2}$ .
Here
$$
\langle K_{S}\rangle=\int_{0}^{\infty} K_{S}(r)|u(r)|^{2}r^{2}\,dr\;,
$$
$$
\langle K_{L}\rangle=\int_{0}^{\infty} K_{L}(r)|u(r)|^{2}r^{2}\,dr\;,
$$
$$
\langle\phi_{SO}r^{2}K_{L}\rangle=
\int_{0}^{\infty}\phi_{SO}(r)r^{2} K_{L}(r)|u(r)|^{2}r^{2}\,dr\;,
$$
$$
K_{S}(r)=\frac{\int_{0}^{r}fg\,dr'}{\int_{0}^{\infty}fg\,dr'}\;,
$$
$$
K_{L}(r)=\frac{\int_{0}^{r}\Bigl(1-\frac{r'^{3}}{r^{3}}\Bigr)
fg\,dr'}{\int_{0}^{\infty}fg\,dr'}\;,
$$
$g$ and $f$ are the radial parts of the Dirac wave function of
the electron,
$u(r)$ is the radial part of the wave function of the odd nucleon.
The functions $K_{S}(r)$ and $K_{L}(r)$ are calculated by
using  simple approximate formulas from [4]
(the relative precision of such a calculation is of order
$\alpha Z R_{0}/(\hbar/mc)$).
Taking $g_{L}=0$ for neutron and $g_{L}=1$ for proton,
we choose
$g_{S}$ to give the experimental value of the magnetic moment
within the single particle approximation:
\begin{eqnarray}
\frac{\mu}{\mu_{N}}=\frac{1}{2}g_{S}+
\Bigl[I-\frac{1}{2}+
\frac{2I+1}{4(I+1)} \frac{m_{p}}{\hbar^{2}}\langle\phi_{SO}r^{2}\rangle
\Bigr]g_{L}
\end{eqnarray}
for $I=L+\frac{1}{2}$ and
\begin{eqnarray}
\frac{\mu}{\mu_{N}}=-\frac{I}{2(I+1)}g_{S}+\Bigl[\frac{I(2I+3)}{2(I+1)}-
\frac{2I+1}{4(I+1)} \frac{m_{p}}{\hbar^{2}}\langle\phi_{SO}r^{2}\rangle
\Bigr]g_{L}
\end{eqnarray}
for $I=L-\frac{1}{2}$.

In the third and fourth columns of
the table 1 we present the values $g_{S}$ and $\varepsilon$
calculated by equations (6)-(9). As one can see from the table,
except $^{159}Tb$ and $^{127}I$, the values $g_{S}$ lie between
the  free Dirac and  free real $g$-factors. For comparison,
in the fifth column
we give the values $\varepsilon$ found in disregarding the
spin-orbit terms in equations (6)-(9) (it corresponds to the calculation
using the original Bohr-Weisskopf formulas [14]). In the last column
we give the values $\varepsilon$ found in [4] by using
a simple, homogeneous over the nucleus, distribution of
$|u(r)|^{2}$. (We note here that in the
case of $^{209}Bi^{82+}$ in [4] a more accurate evaluation of
$\varepsilon$
 was also presented which gave $\varepsilon=0.013$ and
$\lambda=242.0$ nm, without the QED correction.
The same value was  found in [21].)
Taking into account that the single particle
model with the effective $g_{S}$-factor gives
reasonable agreement with experiment for
neutral atoms [22,23], we assume the following
errors bars for $\varepsilon$.
For the ions
where the spin and orbital parts in (6)-(7)
are of the same sign (it results in relatively
large values of $\varepsilon$) the uncertainty
is about 30\% of $\varepsilon$.
For the ions
where the spin and orbital parts in (6)-(7)
are of the opposite sign (it results in relatively
small values of $\varepsilon$) the uncertainty
is about 20\% of
 $\langle K_{S}\rangle$
( $0.2\langle K_{S}\rangle\sim 0.03\alpha Z b$, where
$b$ is a factor tabulated in [4]).
In the case of $Pb$ the uncertainty is assumed
to be about 10\% of $\varepsilon$.
 It should be stressed
however that the uncertainty found in this way
is to be considered only as the order of the expected error.
More accurate calculations of the Bohr-Weisskopf effect
must be based on the many particle nuclear models
and must include a more consequent procedure
for determination of the error bars.

The radiative correction is the sum of the vacuum polarization (VP) and
self energy (SE) contributions. The VP contribution can  easily  be calculated
within the Uehling approximation. We calculated this effect
 for a finite nucleus
charge distribution and found that for $Z$=82, 83 our results
 are in good agreement
with the results of [3,6]. The values of the electric loop
 and  magnetic loop  contributions in the Uehling approximation
 are given
in the second and third columns of the table 2. The Wichman-Kroll (WK)
 contribution
is also the sum of two terms. The first term is given by the WK electric
loop correction to the electron wave function. The calculation of this
term can be done
in the  same way as the calculation of the first order WK contribution.
The results of such a calculation,
based on using the approximate formulas for the WK potential
for a point nucleus [24],
 are given in the
fourth column of the table. The second term is the WK
magnetic loop  contribution. Calculation of
this term is a more complex problem.
However, calculations of the corresponding
 term in the VP screening diagrams for
two-electron ions [25] allow us to expect that this term is small enough.
In the fifth column of the table 2 the total VP contribution, without the
WK magnetic loop  term, is given.
The SE contribution was evaluated in [6,26]
 in a wide interval of Z
 for a finite
nuclear size distribution.
The values of this contribution given in the sixth column of the table 2 are
found by interpolation of the related values from [26].  In the last column
of the table the total radiative correction is listed.

In the table 3 we give the theoretical values of
 the energies and the wavelengths of the transition
between the ground state hyperfine splitting components
of high $Z$ hydrogenlike ions.
The error bars given in the table are mainly
defined by the uncetainty of the Bohr-Wesskopf effect
discussed above.
The magnetic moment values, except $Ho$ [10,11,8], are taken
from[9].
In the case of $Ho$, the theoretical value
constitutes $\lambda=572.5(1.7)$ nm
 and is in good agreement with experiment
$\lambda=572.79(15)$ nm [8].
 The values of
the individual contributions are listed in the table 4.
In the case of $Bi$  the difference between the experimental
value ($\lambda=243.87(2)$ nm [1]) and
 the theoretical value given in the table 3
is within the expected error of the Bohr-Weisskopf effect.
We expect that these theoretical results will be refined
by including the RPA approximation in a more elaborate treatment of
the Bohr-Weisskopf effect [5], based on the dynamic-correlation
model [27]. Such calculations are underway and will be
published elsewhere.

\section*{Acknowledgment}
We thank the authors of  Ref. [8] for making the
results of the paper available to us prior to publication.
 Valuable discussions with S.G.Karshenboim and
S.M.Schneider  are gratefully
acknowledged.
One of us (V.M.S.) wishes to thank the Atomic group
of GSI (Darmstadt) for kind hospitality during a stay
where the major part of this work was done.
The work of V.M.S., A.N.A., and V.A.Y. was supported
in part by Grant No. 95-02-05571a from the Russian Foundation
for Fundamental Investigations.

\begin{table}
\caption {The Bohr-Weisskopf effect within the single particle
model of the nucleus: with taking into account the SO term
in (6)-(9), without the SO term in (6)-(9), and taken from [4].}
\footnotesize
\begin{tabular}{|c|c|c|c|c|l|} \hline
$ion$&Nucleon state&$g_{S}$&$\varepsilon$ (with SO)&$\varepsilon$ (without SO)
&$\varepsilon$ (from [4])\\ \hline
$^{113}In^{48+}$&$1g_{\frac{9}{2}}$&3.67&0.0047&0.0045&0.0039\\ \hline
$^{121}Sb^{50+}$&$2d_{\frac{5}{2}}$&3.04&0.0052&0.0051&0.0046\\ \hline
$^{123}Sb^{50+}$&$1g_{\frac{7}{2}}$&4.21&0.0014&0.0019&0.0016\\ \hline
$^{127}I^{52+}$&$2d_{\frac{5}{2}}$&1.95&0.0052&0.0051&0.0047\\ \hline
$^{133}Cs^{54+}$&$1g_{\frac{7}{2}}$&4.14&0.0017&0.0024&0.0020\\ \hline
$^{139}La^{56+}$&$1g_{\frac{7}{2}}$&3.64&0.0025&0.0032&0.0027\\ \hline
$^{141}Pr^{58+}$&$2d_{\frac{5}{2}}$&4.88&0.0075&0.0073&0.0072\\ \hline
$^{151}Eu^{62+}$&$2d_{\frac{5}{2}}$&3.27&0.0080&0.0079&0.0079\\ \hline
$^{159}Tb^{64+}$&$2d_{\frac{3}{2}}$&-0.203&0.0069&0.0074&0.0073\\ \hline
$^{165}Ho^{66+}$&$1f_{\frac{7}{2}}$&2.90&0.0089&0.0085&0.0086\\ \hline
$^{175}Lu^{70+}$&$1g_{\frac{7}{2}}$&5.10&0.0006&0.0020&0.0018\\ \hline
$^{181}Ta^{72+}$&$1g_{\frac{7}{2}}$&4.76&0.0017&0.0032&0.0030\\ \hline
$^{185}Re^{74+}$&$2d_{\frac{5}{2}}$&2.71&0.0122&0.0120&0.013\\ \hline
$^{203}Tl^{80+}$&$3s_{\frac{1}{2}}$&3.47&0.0179&0.0177&0.020\\ \hline
$^{205}Tl^{80+}$&$3s_{\frac{1}{2}}$&3.50&0.0179&0.0177&0.020\\ \hline
$^{207}Pb^{81+}$&$3p_{\frac{1}{2}}$&-3.56&&0.0419&0.036\\\hline
$^{209}Bi^{82+}$&$1h_{\frac{9}{2}}$&2.80&0.0118&0.0133&0.011\\ \hline
\end{tabular}
\end{table}
\small
\newpage
\begin{table}
\caption{The radiative corrections to the ground state hyperfine
splitting in terms of $x$ defined by equation (1).
 $x_{VP}^{UE}$ is the Uehling electric loop contribution,
 $x_{VP}^{UM}$ is the Uehling magnetic loop  contribution,
 $x_{VP}^{WKE}$ is the WK electric loop  contribution, $x_{VP}$ is
the total VP contribution
without the WK magnetic loop  part, $x_{SE}$ is the SE contribution
found by interpolation of the related values from [26], and
$x_{rad}$ is the total radiative correction.}
\begin{tabular}{|c|c|c|c|c|c|c|} \hline
$Z$&$x_{VP}^{UE}$&$x_{VP}^{UM}$&$x_{VP}^{WKE}$&$x_{VP}$&$x_{SE}$&
$x_{rad}$\\ \hline
49&0.0020&0.0011&-0.0000&0.0031&-0.0074&-0.0043\\ \hline
53&0.0024&0.0013&-0.0000&0.0036&-0.0084&-0.0048\\ \hline
57&0.0029&0.0014&-0.0000&0.0043&-0.0096&-0.0053\\ \hline
63&0.0037&0.0017&-0.0001&0.0054&-0.0116&-0.0062\\ \hline
67&0.0044&0.0020&-0.0001&0.0063&-0.0132&-0.0069\\ \hline
71&0.0053&0.0022&-0.0001&0.0074&-0.0150&-0.0076\\ \hline
75&0.0064&0.0026&-0.0002&0.0088&-0.0171&-0.0083\\ \hline
82&0.0089&0.0033&-0.0003&0.0119&-0.0218&-0.0099\\ \hline
83&0.0094&0.0034&-0.0003&0.0125&-0.0226&-0.0101\\ \hline
\end{tabular}
\end{table}

\newpage
\scriptsize
\begin{table}
\caption{The energies ($\Delta E$) and the wavelengths
($\lambda $) of the transition between
the hyperfine
structure components of the ground state of the hydrogenlike ions.
 $A$ is the
relativistic factor, $\delta$ is the nuclear charge
distribution correction, $\varepsilon$ is the nuclear magnetization
distribution correction (the Bohr-Weisskopf correction), and
$x_{rad}$ is the radiative correction (see equation (1)).}
\begin{tabular}{|c|l|c|c|c|c|c|c|} \hline
$Ion$&$\frac{\mu}{\mu_{N}}$&$A$&$\delta$&
$\varepsilon$&$x_{rad}$&$\Delta E $ (eV)&$\lambda$
(nm)\\ \hline
$^{113}In^{48+}$&5.5289(2)&1.2340&0.0170&0.0047&-0.0043&0.9148(13)
&1355.3(1.9)\\ \hline
$^{121}Sb^{50+}$&3.3634(3)&1.2582&0.0191&0.0052&-0.0045&0.6891(11)
&1799.3(2.8)\\ \hline
$^{123}Sb^{50+}$&2.5498(2)&1.2582&0.0191&0.0014&-0.0045&0.4994(7)
&2482.5(3.5)\\ \hline
$^{127}I^{52+}$&2.81327(8)&1.2843&0.0213&0.0052&-0.0048&0.6587(10)
&1882.2(3.0)\\ \hline
$^{133}Cs^{54+}$&2.58202&1.3125&0.0237&0.0017&-0.0051&0.6582(11)
&1883.6(3.0)\\
\hline
$^{139}La^{56+}$&2.78305&1.3430&0.0263&0.0025&-0.0053&0.8052(14)
&1539.9(2.6)\\ \hline
$^{141}Pr^{58+}$&4.2754(5)&1.3761&0.0292&0.0075&-0.0056&1.464(3)
&847.0(1.9)\\ \hline
$^{151}Eu^{62+}$&3.4717(6)&1.4509&0.0365&0.0080&-0.0062&1.513(4)
&819.4(2.0)\\ \hline
$^{159}Tb^{64+}$&2.014(4)&1.4933&0.0407&0.0069&-0.0065&1.099(3)
&1128(3)\\ \hline
$^{165}Ho^{66+}$&4.132(5)&1.5395&0.0456&0.0089&-0.0069&2.166(7)
&572.5(1.7)\\ \hline
$^{175}Lu^{70+}$&2.2327(11)&1.6453&0.0575&0.0006&-0.0076&1.482(4)
&836.6(2.4)\\
\hline
$^{181}Ta^{72+}$&2.3705(7)&1.7061&0.0645&0.0017&-0.0080&1.758(5)
&705.3(2.2)\\ \hline
$^{185}Re^{74+}$&3.1871(3)&1.7731&0.0706&0.0122&-0.0083&2.749(10)
&451.0(1.7)\\ \hline
$^{203}Tl^{80+}$&1.62226&2.0217&0.0988&0.0179&-0.0096&3.229(18)
&384.0(2.1)\\ \hline
$^{205}Tl^{80+}$&1.63821&2.0217&0.0989&0.0179&-0.0096&3.261(18)
&380.2(2.1)\\ \hline
$^{207}Pb^{81+}$&0.592583(9)&2.0718&0.1049&0.0419&-0.0099&1.215(5)
&1020.5(4.5)\\
\hline
$^{209}Bi^{82+}$&4.1106(2)&2.1250&0.1111&0.0118&-0.0101&5.101(27)
&243.0(1.3)\\ \hline
\end{tabular}
\end{table}
\small
\begin{table}
\caption{ The individual contributions to the ground state hyperfine
splitting in $^{165}Ho^{66+}$
for $\mu=4.132(5)\mu_{N}$ [10,11,8].}
\begin{tabular}{|l|l|} \hline
Nonrelativistic value  &     1.4945(18) eV\\ \hline

Relativistic value (point nucleus) & 2.3007(28) eV\\ \hline

Nuclear size effect &       -0.1050(7)  eV\\ \hline

Bohr-Weisskopf effect  &     -0.0195(59) eV\\ \hline

Vacuum polarization   &      0.0094 eV\\ \hline

Self energy          &      -0.0197 eV\\ \hline

Total theoretical value &    2.1659(66) eV $\;$ [$\lambda$ = 572.5(1.7) nm ]\\ \hline

Experiment [8]  &        2.1645(6)  eV $\;\;$  [$\lambda$ = 572.79(15) nm ]\\ \hline
\end{tabular}
\end{table}

\newpage
\begin{thebibliography}{24}
\bibitem{s1}
I.Klaft, S.Borneis, T.Engel, B.Fricke, R.Grieser, G.Huber,
 T.K\"uhl, D.Marx, R.Neumann, S.Schr\"oder,
P.Seelig, L.V\"olker, Phys.Rev.Lett. {\bf 73}, 2425 (1994).
\bibitem{s2}
M.Finkbeiner, B.Fricke, and T.K\"uhl, Phys.Lett.A {\bf 176}, 113
(1993).
\bibitem{s3}
S.M.Schneider, W.Greiner, and G.Soff, Phys.Rev.A {\bf 50}, 118
(1994).
\bibitem{s4}
V.M.Shabaev, J.Phys.B {\bf 27}, 5825 (1994).
\bibitem{s5}
M.Tomaselli, S.M.Schneider, E.Kankeleit, and T.K\"uhl,
Phys.Rev.C {\bf 51}, 2989 (1995).
\bibitem{s6}
H.Persson, S.M.Schneider, W.Greiner, G.Soff, and I.Lindgren,
Phys.Rev.Lett. {\bf 76}, 1433 (1996).
\bibitem{s7}
H.F.Beyer, H.-J.Kluge, V.P.Shevelko, {\it X-ray Radiation
of Highly Charged Ions} (Springer, Berlin, in press).

\bibitem{s8}
J.R.Crespo Lopez-Urrutia, P.Beiersdorfer, D.Savin,
and K.Widman, Phys. Rev.Lett. {\bf 77}, 826 (1996).
\bibitem{s9}
P.Raghavan, At.Data Nucl.Data Tables {\bf 42}, 189 (1989).
\bibitem{s10}
G.Nachtsheim,
 Pr\"azisionsmessung der Hyperfeinstruktur-Wechselwirkung
von $^{165}Ho$ im Grundzustand,
Ph.D. Thesis, Bonn 1980 (unpublished).
\bibitem{s11}
L.K.Peker, Nucl.Data Sheets {\bf 50}, 137 (1987).
\bibitem{s12}
V.M.Shabaev, M.B.Shabaeva, and I.I.Tupitsyn, Phys.Rev.A {\bf 52},
3686 (1995)
\bibitem{s13}
G.Breit, Phys.Rev. {\bf 35}, 1447 (1930).
\bibitem{s14}
A.Bohr and V.F.Weisskopf, Phys.Rev. {\bf 77}, 94 (1950);
A.Bohr, Phys.Rev. {\bf 81}, 331 (1950);
M. Le Bellac, Nucl.Phys. {\bf 40}, 645 (1963).
\bibitem{s15}
G.Fricke, C.Bernhardt, K.Heilig, L.A.Schaller,
L.Schellenberg,
E.B.Shera, and C.W. de Jager,
At.Data and Nucl.Data Tables {\bf 60},
177 (1995);
H. de Vries, C.W. de Jager, and C. de Vries,
At.Data and Nucl.Data
Tables {\bf 36}, 495 (1987);
W.R.Johnson and G.Soff, At.Data and Nucl.Data Tables {\bf 33}, 405
(1985).
\bibitem{s16}
R.D.Woods and D.S.Saxon, Phys.Rev. {\bf 95}, 577 (1954).
\bibitem{s17}
E.Rost, Phys.Lett.B {\bf 26}, 184 (1968).
\bibitem{s18}
L.A.Sliv and B.A.Volchok, Zh.Eksp.Teor.Fiz. {\bf 36}, 374 (1959)
[Sov.Phys. JETP {\bf 36}, 539 (1959)].
\bibitem{s19}
J.Blomqvist and S.Wahlborn, Ark.f.Fys. {\bf 16}, 545 (1960).
\bibitem{s20}
J.H.D.Jensen and M.G.Mayer, Phys.Rev. {\bf 85}, 1040 (1952).
\bibitem{s21}
L.N.Labzowsky, W.R.Johnson, G.Soff, and S.M.Schneider,
Phys.Rev.A {\bf 51}, 4597 (1995).
\bibitem{s22}
J.T.Eisinger, B.Bederson, and B.T.Feld, Phys.Rev. {\bf 86},
73 (1952).
\bibitem{s23}
H.Kopfermann, {\it Kernmomente} (Akademische Verlagsgesellschaft
M.B.H., Frankfurt am Main, 1956).
\bibitem{s24}
A.G.Fainshtein, N.L.Manakov, and A.A.Nekipelov, J.Phys.B
{\bf 23}, 559 (1990).
\bibitem{s25}
A.N.Artemyev, V.M.Shabaev, and V.A.Yerokhin, Opt.Spectr. (in press).
\bibitem{s26} V.A.Yerokhin, V.M.Shabaev, and A.N.Artemyev,
E-print archive, physics/9705029 (http://xxx.lanl.gov).

\bibitem{s27}
M.Tomaselli, Phys.Rev.C {\bf 37}, 343 (1988); Ann.Phys.(N.Y.) {\bf 205},
362 (1991); Phys.Rev.C {\bf 48}, 2290 (1993).


\end{thebibliography}
\end{document}